\documentclass[a4paper,11pt]{article} 
\usepackage{graphicx}
\usepackage{url}
\usepackage{latexsym}
\newtheorem{theorem}{Theorem}

\title{Interaction Maxima in Distributed Systems}

\author{Thomas Robertazzi\thanks{ECE Dept., Stony Brook University, NY 11794,
Stony Brook, New York, United States} 
\and 
Maciej Drozdowski\thanks{Institute of Computing Science, 
Pozna\'n University of Technology, 
Piotrowo 2, 60-965 Pozna\'n, Poland.
orcid: orcid.org/0000-0001-9314-529X}} 
\date{}

\begin{document}
\maketitle

\begin{abstract}
In this paper we study the maximum degree of interaction which may emerge 
in distributed systems.
It is assumed that a distributed system is represented by a graph of 
nodes interacting over edges.
Each node has some amount of data.
The intensity of interaction over an edge is proportional to the product 
of the amounts of data in each node at either end of the edge.
The maximum sum of interactions over the edges is searched for.
This model can be extended to other interacting entities.
For bipartite graphs and odd-length cycles we prove that the greatest 
degree of interaction emerge when the whole data is concentrated 
in an arbitrary pair of neighbors.
Equal partitioning of the load is shown to be optimum for complete graphs.
Finally, we show that in general graphs for maximum interaction
the data should be distributed equally between the nodes 
of the largest clique in the graph.
We also present in this context a result of 
Motzkin and Straus from 1965 for the maximal interaction objective.
\end{abstract}

{\bf Keywords:}
distributed systems,
interaction maxima,
quadratic programming,
maximum clique.

\section{Introduction}

As technology advances, the amount of data humanity 
generates, stores and processes is increasing at rapid rates.  
One projection is for a global data sphere of 175 zettabytes 
(175 x 10$^{21}$ bytes) by 2025 \cite{Ca20}.  
Global mobile data traffic according to another projection 
\cite{stat20} 
is expected to be 77 exabytes per month in 2022. 
Hence, means are needed for understanding the operation and interaction of 
communicating entities and data storage in networked systems in this 
context.  
Mathematical performance evaluation offers a method to accomplish 
this and had much success in the past and current environment going back to 
Erlang's sizing of telephone systems using queuing theory.  
Some applications appear natural for the mathematical model 
discussed in this paper.

In the first application consider networked systems where a network is 
represented mathematically by a graph and the $i$th node contains a memory 
of size $m_i$.  
It is useful to asses the total amount of communication in order
to size the communication infrastructure for this traffic.
We make an assumption that two adjacent nodes  interact and have 
an interaction activity proportional to the product of the sizes of their 
memories, $m_im_j$ for the $i$th and $j$th node.  
More specifically, the two-way traffic on the link connecting the two 
adjacent memories is $cm_im_j$ where $m_i$ have units of [bits]
and $c$ is a proportionality constant with 
units of [1/[bits$\times$sec]].
This assumption makes some sense if one thinks that the larger a node's 
memory the more likely is its node to query other nodes and be queried by 
other nodes.  
There is an implicit assumption here that nodal memories are fully utilized 
(a very large memory with no data would not show much activity).   
Overall, in a sense we are considering an intuitive model of a large 
database system emphasizing its interacting component memories.

The second application involves telecommunications between groups of 
communicating entities (people and/or machines) modeled as nodes
(see a similar idea in \cite{Met13}).  
The communication could be implemented physically in terms of a wired 
Internet or by cellular wireless communications.  
Using the same mathematical model as above, the network is represented by a 
graph where the $i$th node contains a population of communicating entities 
of size $m_i$.  
Since members of the population both produce and receive information,
we assume that two adjacent nodes' populations interact and have 
a communication intensity proportional to the product of the size of their 
populations $m_im_j$ for the $i$th and $j$th nodes.  
Again, the two-way traffic on the link connecting the two adjacent 
populations is $cm_im_j$ where $c$ is a proportionality constant 
with units of [bits/(entities$^2\times$ sec)] and the $m_i$ have units of 
[entities]. 
It is practical to assess total communication flow created by 
the interacting entities to estimate their stress on the communication system
and verify its suitability for such kind of interaction.
The second model also makes some sense if one believes that the larger a 
node's population of communicating entities, the more likely the node is to 
have communication with entities in other nodes.  
There is an implicit assumption here that each nodal population has the same 
level (percentage) of activity as that of other nodes. 
These are reasonable assumptions for a first study of this problem.   

These two examples do not preclude other application areas.  
Such product models arise in physical systems involving charge or gravity, 
e.g., in N-body simulations.
Intensity of interaction in scatter/gather and gossiping communications
in computer networks can be modeled similarly.
In the social sciences and economics one could see the nodal quantities, 
being influence or money, mathematically along the same lines as above.
In this context there are interesting optimization problems 
which are presented below.

\section{Problem formulation}
\label{sec:formul}

For conciseness of presentation we will be referring to memory/population 
size in some node as the node {\em load}. 
We will refer to {\em flow} as the amount of interaction.
The problem considered in this paper is formulated as follows:

\smallskip

\noindent
{\bf Data:}
A connected graph $G=(V,E)$ with $n$ vertices (nodes) is given.
Set $E$ of edges is defined by connectivity matrix:
$l_{ij}=1$ if nodes $i$ and $j$ are connected, and 0 otherwise,
for $i,j=1,\dots,n$.
Vertex loads (memory/population) sizes are $m_j\geq 0, j=1,\dots,n$.
Loads of all vertices can be represented as a vector
$\vec{m}=[m_1,\dots,m_n]^T$.
Let $D$ denote the total load volume in $G$, and let
edge flows be defined as:
$f_{ij}=cm_im_j$ for $i,j=1,\dots,n$ where 
$c$ is the flow factor used for a conversion to bits per second.
For brevity of exposition, we will we will assume that $c=1$.

\medskip

\noindent
{\bf Problem:}
\begin{eqnarray}
\max_{\vec{m}=[m_1,\dots,m_n]}{\cal F}(\vec{m})=
\sum_{i=1}^n\sum_{j=1}^nf_{ij}=
\sum_{i=1}^n\sum_{j=1}^ncl_{ij}m_im_j\label{eq:obj}&&\\
\textrm{s.t.:}~~~~\sum_{i=1}^nm_i=D\label{eq:2nd}%
\qquad \qquad \qquad&&\\
m_i\geq 0\quad i=1,\dots,n\label{eq:3rd}&&
\end{eqnarray}

The goal here is to maximize the sum of the edge flows.
This value provides useful information on sizing bandwidth of communication 
links over which the interaction is conducted, as maximum of 
${\cal F}(\vec{m})$ is the worst case (upper bound) for the flow.  
The problem is non-trivial because of the constraints 
(\ref{eq:2nd}) and (\ref{eq:3rd}).
More specifically, the problem we discuss is a quadratic optimization 
\cite{G20} problem with inequality constraints.  
There are numerous applications of such problems to areas such as portfolio 
optimization, signal and image processing, least squares approximation and estimation and control 
theory.   
There is much work on the numerical solution of quadratic 
optimization problems.   
Numerical techniques that can be applied to this problem include active set 
and interior point methods.  
The website at \cite{GT20} lists approximately thirty different solution codes.
However, in this paper we are interested more in what can be found 
analytically. 
A quadratic programming formulation with inequalities  
is an {\bf NP}-hard problem \cite{GJ79} in general.

%
%

\section{Single-Level Tree}
\label{sec-slt}

Let 1 be the index of the root node, and $2,\dots,n$ the indices 
of the leaf nodes.
By using (\ref{eq:2nd}) problem (\ref{eq:obj})-(\ref{eq:3rd}) 
can be re-stated as follows:
\begin{eqnarray}
\max_{m_1,\dots,m_n}~~{\cal F}(\vec{m})=~\sum_{j=2}^nf_{1j}=&&\\
\sum_{j=2}^nm_1m_j=m_1\sum_{j=2}^nm_j=m_1(D-m_1)&&\label{e:SLtree-1}
\end{eqnarray}
Thus, for a single-level tree ${\cal F}(\vec{m})$ is actually
${\cal F}(m_1)$, i.e. a function of one variable $m_1$.
By (\ref{e:SLtree-1}) ${\cal F}(m_1)$ has maximum of $D^2/4$ at 
$m_1=D/2, \sum_{i=2}^nm_i=D/2$.
It is immaterial how the load is split between leaf
nodes as long as together they have load $D/2$ in total.
For  maximum of ${\cal F}(\vec{m})$ the load equally well can be split 
in halves between the root and one leaf node.

%
%

\section{Bipartite Graphs}
\label{sec:bgr}

In this section we show that any load distribution in a bipartite graph
can be reduced to the single-level tree distribution
without decreasing the flow objective (\ref{eq:obj}).
Actually, a bipartite graph can be reduced to a special case of
a single-level tree: to just one edge.
In this sense we say that bipartite graphs are equivalent 
to single-level trees.
Note that bipartite graphs include trees, even-length cycles, 
and more generally, all graphs without odd-length cycles.
Assume a bipartite graph $G(X,Y,E)$ is given, where $X, Y$ are the two
parts of vertex set and $E$ is the set of edges.

\begin{theorem}
\label{theo:bipartite}
Any load distribution in a bipartite graph can be reduced
to a single-level tree distribution without decreasing
the flow objective (\ref{eq:obj}).
\end{theorem}

\noindent
{\bf Proof.}
Consider a bipartite graph $G(X,Y,E)$.
Let $\{x,y\}\in E$ denote an edge.
Without loss of generality $x\in X, y\in Y$.
The flow in $G$ is
\begin{eqnarray}
{\cal F}=\sum_{\{x,y\}\in E}cm_xm_y\leq
\sum_{x\in X}cm_x\left(\sum_{y\in Y}m_y\right)=
c\left(\sum_{x\in X}m_x\right)\left(\sum_{y\in Y}m_y\right)\label{eq:bipa-nodec}
\end{eqnarray}
Thus, by inequality (\ref{eq:bipa-nodec}) load of all vertices in set $X$
can be moved to a single super-vertex $x'$ with load 
$\sum_{x\in X}m_x$ and of all vertices in set $Y$ to a single
super-vertex $y'$ with load $\sum_{y\in Y}m_y$
connected by one edge without decreasing objective ${\cal F}$.
As the super-vertices $x',y'$ any vertices can be chosen 
such that $x'\in X, y'\in Y, \{x',y'\}\in E$.
\hfill $\Box$

\medskip

Note that proof of Theorem~\ref{theo:bipartite} defines a load 
distribution transformation from any distribution in a bipartite graph 
to a load distribution on a single edge.
In this load distribution transformation flow ${\cal F}$
is nondecreasing.
But this transformation does not immediately construct optimal 
load partitioning.
For optimality of the partitioning the load size in super-vertex $x'$
must be equal to the load size in super-vertex $y'$ as shown in 
Section~\ref{sec-slt}.

%
%

\section{Odd-Length Cycles}
\label{sec:olc}

We will show that any load distribution in an odd-length cycle with 
$n\geq 5$ nodes can be reduced to a load distribution in a single-level 
tree, or even, a distribution over an edge.

\begin{theorem}
\label{theo:odd-len-cyc}
Any load distribution in an odd-length cycle of length at least 5
can be reduced to a single-level tree distribution without decreasing
the flow objective (\ref{eq:obj}).
\end{theorem}

\noindent
{\bf Proof.}
Suppose that the load distribution is equal, i.e. $m_1=\dots=m_n=D/n$.
The total flow for equal load distribution is
${\cal F}_{eq}=n\times(D/n)^2=D^2/n$.
However, equal distribution of the load between two neighboring nodes
results in flow ${\cal F}'=(D/2)\times(D/2)=D^2/4$.
Flow value ${\cal F}'$ is bigger than ${\cal F}_{eq}$ for all $n\geq 5$.
Thus, equal load partitioning in odd-size cycles can be transformed
to equal load distribution over a single edge without decreasing 
the value of the flow objective.

\begin{figure}[t]
\setlength{\unitlength}{1cm}
\begin{picture}(13,3)
\put(0,0.2){\includegraphics[width=\textwidth]{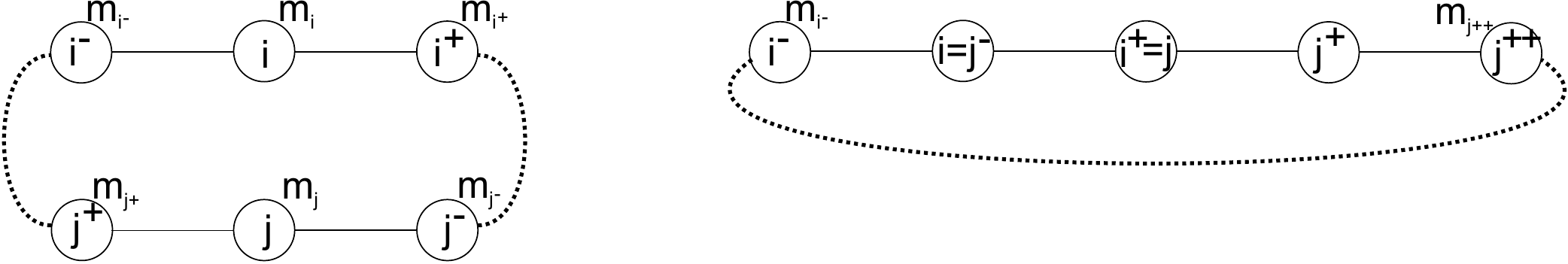}}
\put(2,-0.4){a)}
\put(6,-0.4){b)}
\end{picture} 
\caption{Node neighborhood overlap in the proof of
Theorem~\ref{theo:odd-len-cyc}.
a) no overlap, b) 2-node overlap.}
\label{fig-4theo-olc}
\end{figure}

Suppose that the load distribution is not equal, i.e. $m_i$
are not all simultaneously equal to the same value.
Consider some node $i$ and its neighbors with indices
$i^-=(i-1)\bmod n$ and $i^+=(i+1)\bmod n$.
Let $\phi(i)$ denote the sum of loads in neighbors of node $i$.
Node $i$ together with its neighbors contribute to the flow
value $m_i(m_{i^-}+m_{i^+})=m_i\phi(i)$.
Since load distribution is not equal, nodes in the cycle can be ordered
according to the increasing values of $\phi$.
Let us assume that node $i$ has the smallest $\phi(i)$
and node $j$ has the largest $\phi(j)$ in the cycle.
If the set of $i$ and $j$ neighbors are disjoint
(Fig.\ref{fig-4theo-olc}a), i.e.
$\{i^{-},i,i^{+}\}\cap\{j^{-},j,j^{+}\}=\emptyset$
then by shifting the whole load of $m_i$ to node $j$ we
lose flow value $m_i\phi(i)=m_i(m_{i^-}+m_{i^+})$,
but we gain $m_i(m_{j^-}+m_{j^+})=m_i\phi(j)$.
Overall we gain in this load shift because load distribution is unequal
and $\phi(i)$ is the smallest in the cycle.

Suppose $|\{i^{-},i,i^{+}\}\cap\{j^{-},j,j^{+}\}|=1$,
that is the set of $i$ and $j$ neighbors overlap in one node.
But since $\phi(i)\leq \phi(j)$, we do not lose on the value of flow
by shifting whole load $m_i$ to $j$.
Finally, suppose $|\{i^{-},i,i^{+}\}\cap\{j^{-},j,j^{+}\}|=2$ 
(Fig.\ref{fig-4theo-olc}b) 
which means that $i$ is a direct neighbor of $j$ and vice versa.
Without loss of generality let us assume that $j^{+}$ is not a neighbor 
of $i$ and $i^{-}$ is not a neighbor of $j$.
Let $j^{++}$ be the neighbor of $j^{+}$ which is not $j$.
Node $j^{++}$ exists because $n\geq 5$.
By shifting whole load $m_i$ to $j^{+}$ we lose flow value
$m_i\phi(i)=m_i(m_{i^-}+m_{i^+})=
m_i(m_{i^-}+m_j)$,
but we gain flow
$m_i(m_j+m_{j^{++}})=m_i\phi(j^{+})$.
Overall, we gain as a result of this load shift because
$\phi(i)\leq\phi(j^{++})$ (because $\phi(i)$ is the smallest in the cycle).
The three cases of shifting load from $i$ with the smallest
$\phi(i)$ result in load distribution with new $m_i=0$.
Hence, the load distribution is as in a chain,
which is a bipartite graph.
The load distribution can be further transformed to a single-level tree
distribution as in Theorem~\ref{theo:bipartite} without decreasing
value of flow ${\cal F}$.
\hfill $\Box$

%
%

\section{Clique}
\label{sec:cliq}

We will show by induction that equal partitioning is the optimum 
distribution of the load in a clique (a complete graph).

\begin{theorem}
\label{theo:clique-n}
In a clique of size $n$ the flow is maximum if $m_i=D/n$, for $i=1,\dots,n$
and the value of the flow is
${\cal F}=(D/n)^2\times\left( n \atop 2\right)$.
\end{theorem}

\noindent
{\bf Proof.}
Note that this theorem holds for $n=2$ (one edge) because 
$\left( 2 \atop 2\right)=1$ and by the result in Section \ref{sec-slt}.
Let $OPT(x,y)$ denote the optimum value of the flow 
in a clique with $x$ nodes and volume of load $y$.
Suppose that the theorem holds for $n\geq 2$ then it also holds for $n+1$.
The flow is
\begin{eqnarray}
{\cal F}=m_{n+1}\sum_{i=1}^{n}m_i+OPT(n,D-m_{n+1})&=&\\
m_{n+1}(D-m_{n+1})+OPT(n,D-m_{n+1})\label{eq-cliq-n}.
\end{eqnarray}
The first component in (\ref{eq-cliq-n}) corresponds to the flow between 
node $n+1$ and the remaining part of the clique.
The second component is the flow for the optimum distribution of the load 
of the remaining size $D-m_{n+1}$ in a clique with $n$ nodes.
By the assumption of the inductive proof 
\begin{equation}
OPT(n,D-m_{n+1})=
\frac{(D-m_{n+1})^2}{n^2}\times
\left(
\begin{array}{c}
n\\
2
\end{array}
\right)\label{eq-opt-cliq-n}
\end{equation}
Let 
\begin{eqnarray}
\lambda=
\frac{
\left(
\begin{array}{c}
n\\
2
\end{array}
\right)}
{n^2}=
\frac{n!}{2!(n-2)!n^2}=
\frac{(n-1)n}{2!n^2}=\frac{n-1}{2n}\label{eq-what-is-lambda}
\end{eqnarray}
Equation (\ref{eq-cliq-n}) can be rewritten as:
\begin{eqnarray}
{\cal F}=
m_{n+1}(D-m_{n+1})
+(D-m_{n+1})^2\lambda&=&\\
-(1-\lambda)m_{n+1}^2+
D(1-2\lambda)m_{n+1}+D^2\lambda.
\end{eqnarray}
This is a quadratic function of $m_{n+1}$ with a maximum at
\begin{eqnarray}
m^*_n=
\frac{1-2\lambda}{2-2\lambda}D=
\frac{1-\frac{2(n-1)}{2n}}{2-\frac{2(n-1)}{2n}}D=&&\\
\frac{n-(n-1)}{2n-(n-1)}D=
\frac{D}{n+1}
\end{eqnarray}
The load assignments of the other nodes, by the assumption 
of the inductive proof are
\begin{eqnarray}
m^*_i=
\frac{D-D/(n+1)}{n}=
\frac{(n+1)D-D}{n(n+1)}=
\frac{D}{n+1}
\end{eqnarray}
By (\ref{eq-cliq-n}), (\ref{eq-opt-cliq-n}), (\ref{eq-what-is-lambda}) 
the optimum value of the objective function is:
\begin{eqnarray}
{\cal F}=
\frac{D}{n+1}\frac{nD}{n+1}+
\frac{D^2n^2}{(n+1)^2}\frac{(n-1)}{2n}=
\frac{D^2}{(n+1)^2}
\left[
n+\frac{n(n-1)}{2}
\right]=&&\\
\frac{D^2}{(n+1)^2}
\left[
\frac{n(n+1)}{2}
\right]=
\frac{D^2}{(n+1)^2}
\left(
\begin{array}{c}
n+1\\
2
\end{array}
\right)
&&
\end{eqnarray}
Thus, from the assumption that the theorem holds for $n\geq 2$
it follows that it holds also for any $n$.
\hfill$\Box$

%
%

\section{General graphs}
\label{sec:gengra}

In Sections \ref{sec:bgr}, \ref{sec:olc} we have shown a set of load 
distribution transformations reducing the size (i.e. order) of 
the subgraph with non-zero node loads without decreasing value 
of the flow ${\cal F}$.
In this section we generalize this approach and analyze 
the smallest subgraph which allows the value of flow ${\cal F}$
that is not smaller (but may be larger) than the initial one.
Thus, if the initial load distribution was optimum with respect to 
${\cal F}$, then also after the transformations we propose,
the value of the flow remains optimum.
Consequently, characterizing such a smallest subgraph also determines
the optimum solutions.
We will show that the smallest subgraph remaining after a sequence
of flow-preserving transformations is the largest clique in the graph.
This will be achieved by generalizing the transformation
used in the proof of Theorem \ref{theo:odd-len-cyc}.

\begin{theorem}
\label{theo:gengra}
Flow ${\cal F}$ is maximum if the load is distributed equally between 
nodes of the largest clique in the graph.
\end{theorem}

\noindent
{\bf Proof.}
Consider two nodes $i, j$ of graph $G$ and their direct 
neighborhoods $N(i), N(j)$, respectively.
Direct neighborhood of some nodes $a,b$ means that there is edge
$\{a,b\}\in E$.
Let $\phi(a)$ denote the sum of the loads in neighbors of some node $a$.
The contributions of nodes $i, j$ to the value of flow are 
$m_i\phi(i), m_j\phi(j)$, respectively.
Suppose $i\not\in N(j), j\not\in N(i)$.
Without loss of generality we can assume that $\phi(i)\leq\phi(j)$.
Then, it is possible to transfer the whole load $m_i$ to node $j$
without decreasing the value of flow ${\cal F}$ 
(the value of flow may increase).
As an additional effect of this transformation the size
of the subgraph with nodes $i$ holding $m_i>0$ decreases.
This procedure may be continued as long as there are nodes $i, j$ 
satisfying $i\not\in N(j), j\not\in N(i)$.
The procedure stops when all pairs of nodes are direct neighbors.
Consequently, the load will be collected in one or more disconnected 
cliques in $G$.
By the lack of connectivity we mean here that there is no path
connecting the cliques over nodes $l$ with $m_l>0$.
Suppose there are two cliques $K_a$ and $K_b$ in $G$ remaining after
the above sequence of load shifts and they hold loads $x,y\geq 0$, 
respectively.
$K_a$ has size $r$, $K_b$ has size $s$, and $r\geq s$.
According to Theorem \ref{theo:clique-n} the maximum flows in $K_a$
is $x^2/r^2\times({r \atop 2})=x^2(1-\frac{1}{r})/2$.
Analogously, the maximum flows in $K_b$ is $y^2(1-\frac{1}{s})/2$.
Collecting loads $x$ and $y$ in the larger clique $K_a$
results in flow 
$(x+y)^2(1-\frac{1}{r})/2\geq x^2(1-\frac{1}{r})/2$,
$(x+y)^2(1-\frac{1}{r})/2\geq y^2(1-\frac{1}{s})/2$
and
$(x+y)^2(1-\frac{1}{r})/2\geq (x+y)^2(1-\frac{1}{s})/2$.
Hence, concentrating the load in the largest clique in $G$ 
results in the optimum flow ${\cal F}$.
\hfill$\Box$

\bigskip

Let us observe that in the above proof we exploited property
$\phi(i)\leq \phi(j)\Rightarrow 
m_i\phi(i)+m_j\phi(j)\leq (m_i+m_j)\phi(j)$ of the flow function.
Thus, the above proof can be extended to other forms of the flow
objective which are additive functions of nodal flows $f$
with the property:
$\phi(i)\leq \phi(j)\Rightarrow 
f(m_i,\phi(i))+f(m_j,\phi(j))\leq
f(m_i+m_j,\phi(j))$.

%
%
\section{Motzkin and Straus result}
\label{sec:MSres}

In 1965 article \cite{MS65} T. S. Motzkin and E. G. Straus 
solved a problem suggested in 1963 by J.E.MacDonald \cite{McD63}.
Their result, expressed in terms of our paper, states that
given weights $m_1,\dots,m_n$ on $n$ vertices of graph $G$,
function ${\cal F}$ in equation (\ref{eq:obj}) has maximum 
$\frac{D^2}{2}(1-\frac{1}{\omega(G)})$
when node loads are $m_1=\dots=m_n=D/\omega(G)$,
where $\omega(G)$ is the size of the largest clique in $G$.
Their proof used the idea of maximization of flow ${\cal F}$
over an $n$-dimensional simplex on nodes of $G$.
Our proof of Theorem \ref{theo:gengra} provides an analogous result
albeit on a different premise.
Note that both Motzkin and Straus result and our Theorem \ref{theo:gengra} 
hinge on finding largest clique in a graph which is {\bf NP}-hard 
\cite{GJ79}, but still, has many applications \cite{DDNA20}.

\section{Conclusions}
\label{sec-concl}

It is both timely and very interesting to see what is and isn't possible 
with the use of $m_i m_j$ modeling of interaction in this problem.  
We have related such modeling to:

\noindent
$\bullet$ a large number of application areas that are likely to be 
important both in the future as well as in the present,

\noindent
$\bullet$ demonstrating simple solutions for maximal flow in single-level 
tree, bipartite graphs and clique interconnection networks,

\noindent
$\bullet$ our Theorem \ref{theo:gengra} and the Motzkin-Straus result 
provide an elegant graph theoretic solution for this problem.

Furthermore, if the gain in forming a social group can be modeled 
as in our multiplicative model, then social implications, 
e.g. in emerging of communities, can be considered.
The load, data, power, money can be concentrated in the largest 
clique $K$, a tightly connected subgraph of $G$,
for maximum flow ${\cal F}$.
The remaining subgraph $G\setminus K$ is immaterial.
Still, there may be practical reasons for using a non-optimal flow,
such as resiliency.
Thus, the multiplicative interaction model is an intriguing one which could 
be of good utility in future and current applications.

\section*{Acknowledgments}
We thank 
Ji Liu,
Heather Macbeth,
Janet MacDonald
and
Yue Zhao for useful discussions on this problem.
This research did not receive any specific grant from funding agencies 
in the public, commercial, or not-for-profit sectors.


\begin{thebibliography}{22}

\bibitem{Ca20}
M.Campbell, 
DNA Data Storage: Automated DNA Synthesis and Sequencing are Key to 
Unlocking Virtually Unlimited Data Storage, Computer 53 (2020) 63--67. 
doi: 10.1109/MC.2020.2967908.

\bibitem{DDNA20}
A. Douik, H. Dahrouj, T.Y. Al-Naffouri and M.-S. Alouini, 
A Tutorial on Clique Problems in Communications and Signal Processing, 
Proc. of the IEEE 108 (2020) 583--608.
doi: 10.1109/JPROC.2020.2977595.

\bibitem{GJ79}
M.~R. Garey and D.~S. Johnson.
Computers and Intractability, a Guide to the Theory of NP-Completeness.
W.H. Freeman, 1979.

\bibitem{G20}
A. Geletu,
Quadratic programming problems - a review on algorithms and applications 
(Active-set and interior point methods),
slides, Ilmenau University of Technology.
\url{https://www.tu-ilmenau.de/fileadmin/media/simulation/Lehre/Vorlesungsskripte/Lecture_materials_Abebe/QPs_with_IPM_and_ASM.pdf}
(accessed Feb 18, 2021).

\bibitem{GT20}
N. Gould and P. Toint, 
A Quadratic Programming Page. 
\url{http://www.numerical.rl.ac.uk/qp/qp.html}
(accessed Feb 18, 2021).

\bibitem{McD63}
J.E.MacDonald Jr., Problem E1643, Amer. Math. Monthly, 70 (1963), 1099

\bibitem{Met13}
Bob Metcalfe,
Metcalfe's Law after 40 Years of Ethernet,
Computer 46 (2013) 26--31.
doi: 10.1109/MC.2013.374

\bibitem{MS65}
T.S.Motzkin, E.G.Straus,
Maxima for graphs and a new proof of a theorem of Tur\'{a}n
Canadian J. of Math., 17, pp. 533-540, 1965
doi: 10.4153/CJM-1965-053-6

\bibitem{stat20}
Statistica web page. \url{https://www.statistica.com}
(accessed July 9, 2020).

\end{thebibliography}
\end{document}